\documentclass[reqno]{amsart}
\usepackage[latin1]{inputenc}
\usepackage{amsmath}
\usepackage{amsfonts}
\usepackage{amssymb}

\def\Re{ Re }

\def\Aa{A^-}
\def\Ac{A^+}
\def\1{{\mathbf{1}}}
\def\chL{\kappa}

\def\H{\mathcal{H}}
\def\Hf{H_f}
\def\infspec{{\rm inf \, spec \, }}

\def\Pf{P_f}
\def\PFop{H}  

\def\C{{\mathbb{C}}} 
\def\N{{\mathbb{N}}} 
\def\R{{\mathbb{R}}}  

\def\vac{\Omega_f}

\newcommand{\la}{\langle}
\newcommand{\ra}{\rangle}

\def\bra{\big\langle}

\def\ket{\big\rangle}

\def\Fo{{\mathfrak F}}
\def\H{{\mathcal H}}

\def\eqnn{\begin{eqnarray*}}
\def\eeqnn{\end{eqnarray*}}
\def\eqn{\begin{eqnarray}}
\def\eeqn{\end{eqnarray}}

\newtheorem{theorem}{Theorem}[section]

\newtheorem{proposition}{Proposition}[section]

\newtheorem{remark}{Remark}[section]

\begin{document}

\title[Ground state energy of the Pauli-Fierz model]
{On the ground state energy of the translation invariant Pauli-Fierz model}

\author[J.-M. Barbaroux]{Jean-Marie Barbaroux}

\address{J.-M. Barbaroux, Centre de Physique Th\'eorique, Luminy Case 907, 13288
Marseille Cedex~9, France,  and D\'epartement de Math\'ematiques,
Universit\'e du Sud Toulon-Var, 83957 La Garde Cedex, France}
\email{barbarou@univ-tln.fr}

\author[T. Chen]{Thomas Chen}

\address{T. Chen, Department of Mathematics, Princeton University,
Fine Hall, Washington Road, Princeton, NJ 08544, U.S.A.}
\email{tc@math.princeton.edu}

\author[V. Vougalter]{Vitali Vougalter}

\address{V. Vougalter, Mathematics Department,
202 Mathematical Sciences Bldg,
University of Missouri,
Columbia, MO 65211-4100 USA}
\email{vitali@math.missouri.edu}

\author[S. Vugalter]{Semjon Vugalter}

\address{S. A. Vugalter, Mathematisches Institut, Ludwig-Maximilians-Universit\"at
M\"unchen, Theresienstrasse 39, 80333 M\"unchen, and Institut f\"ur
Analysis, Dynamik und Modellierung, Universit\"at Stuttgart}
\email{wugalter@mathematik.uni-muenchen.de}

\begin{abstract}
In this note, we determine the ground state energy of the translation
invariant Pauli-Fierz model to subleading order $O(\alpha^3)$ with respect to powers
of the finestructure constant $\alpha$, and prove rigorous error bounds
of order $O(\alpha^{4})$. A main objective of our argument is its brevity.
\end{abstract}

\maketitle

\section{Introduction}

We study the translation invariant Pauli-Fierz model describing a spinless electron
interacting with the quantized electromagnetic radiation field.
We present a very short and simple method to determine the subleading terms
of the ground state energy up to
order $O(\alpha^3)$, where $0<\alpha\ll1$ denotes the finestructure constant,
and to rigorously bound the error by $O(\alpha^4)$.

A well-known difficulty connected to this problem arises from the fact that
the ground state energy is not an isolated eigenvalue of the Hamiltonian,
and that the form factor in the interaction
term of the Hamiltonian contains a critical frequency space singularity
(the {\em infrared problem} of Quantum Electrodynamics (QED)).
  One of the most striking consequences is that the
  ground state energy does not exist as a convergent power series in $\alpha$.
  It is inaccessible to ordinary perturbation theory,
  and can only be written as a convergent expansion in powers of $\alpha$
  of the form $\sum_{n} b_n(\alpha) \, \alpha^n$, with $\alpha$-dependent coefficients
  $b_n(\alpha)$ that {\em diverge} in the limit $\alpha\rightarrow0$ at a
  sub-power rate, i.e.,
  $|b_n(\alpha)|=o(\alpha^{-\delta} )$ for any $\delta>0$, \cite{BCFS2,BFP,Chen2006}.

  It is possible to
  determine the ground state energy and the renormalized electron
  mass to any arbitrary precision in powers of $\alpha$, with rigorous error bounds,
  by use of sophisticated rigorous renormalization group methods,
  \cite{BCFS2,BFP,Chen2006}.
  However, these algorithms are highly complex, and explicitly
  computing the ground state energy to any subleading order
  in powers of $\alpha$ is a
  voluminous task.

  Therefore, it remains a desirable goal to devise alternative methods
  that produce such results more directly for non-trivial, but reasonably low
  orders in powers of $\alpha$ with rigorous error bounds, and with much less effort.

The present paper is the first in a number of works
investigating an approach to such problems based on perturbations
around the {\em true ground state} of the interacting system.
A key ingredient in our argument is a new estimate on the expected photon
number in the ground state derived from \cite{ChFr2006}
whose proof involves a bound on
the renormalized electron mass uniform in the infrared cutoff, \cite{BCFS2,Chen2006}.

  Estimates on the ground state energy play an important role, for instance, in
  binding problems, e.g., the determination of the hydrogen binding energy.
  The leading term of the ground state energy was demonstrated to be of order $O(\alpha^2)$ (up to normal
  ordering) in \cite{HVV}, and explicitly determined, with a rigorous error bound of order
  $O(\alpha^3)$.
  A similar result was subsequently obtained for the spin $\frac12$ case in \cite{CattoHainzl2004}.
  In \cite{HaHiSp2005}, the ground state energy for the Bogoliubov-transformed,
  translation-invariant Nelson model is explicitly determined up to $O(\alpha^3)$, with
  an error bound of order $O(\alpha^{\frac72})$ (uniformly in the ultraviolet cutoff).
  While for a fixed ultraviolet cutoff, this result is 
  fully analogous to ours, we present here a new and 
  particularly short method, and our sharp error bound of order $O(\alpha^4)$ is necessary
  for a companion paper, where we address the hydrogen ground state energy.

\section{Definition of the model}
\label{sec:model}



We study a non-relativistic electron interacting with the
quantized electromagnetic field in Coulomb gauge.
The Hilbert space accounting for the pure states of the
electron is given by $L^2(\R^3)$, where we neglect its spin.
The Fock space of transverse photon states in the Coulomb gauge is given by
 $$
   \Fo \; = \; \bigoplus_{n \in \N} \Fo_n ,
 $$
where the n-photon space $\Fo_n =
\bigotimes_s^n\left(L^2(\R^3)\otimes\C^2\right)$ is the symmetric
tensor product of $n$ copies of one-photon Hilbert spaces $L^2(\R^3)\otimes\C^2$.
The factor $\C^2$ accounts
for the two independent transversal polarizations of the photon.
On $\Fo$, we introduce creation- and annihilation operators
$a_\lambda^*(k)$, $a_\lambda(k)$ satisfying the distributional commutation
relations
 $$
  [ \, a_{\lambda}(k) \, , \, a^\ast_{\lambda'}(k') \, ] \; = \;
  \delta_{\lambda, \lambda'} \, \delta (k-k')
  \; \;   ,
  \quad [ \, a_\lambda^\sharp(k) \, , \, a_{\lambda'}^\sharp(k') \, ] \; = \; 0 ,
 $$
where $a^\sharp_\lambda$ denotes either $a_\lambda$ or $a_\lambda^*$.
There exists a unique unit ray $\vac\in\Fo$, the Fock vacuum,
which satisfies $a_\lambda(k) \, \vac=0$ for all $k\in\R^3$ and
$\lambda\in\{+,-\}$.

The Hilbert space of states of the system consisting of both the electron
and the radiation field is given by
$$
    \H \; := \; L^2(\R^3) \, \otimes \, \Fo .
$$
We use units such that $\hbar = c = 1$, and where the mass of the
electron equals $m=\frac12$. The electron charge is then given by
$e=\sqrt{\alpha}$, where the finestructure
constant $\alpha$ will here be considered as a small parameter.

The Hamiltonian of the system is given by
\begin{equation}\label{rpf}
 \PFop  \; = \;  \1_{el} \, \otimes \, \Hf \, + \, : \, \left( \, i\nabla_{x} \, \otimes \, \1_f \, - \,
 \sqrt{\alpha} \, A(x) \, \right)^2 \, :
\end{equation}
where $: \, ( \, \cdots \, ) \, :$ denotes normal ordering
(corresponding to the subtraction of a normal ordering constant proportional to $\alpha$),
\begin{equation}\nonumber
 \Hf \; = \; \sum_{\lambda= +,-} \int_{\R^3} dk \, |k| \, a_\lambda^\ast (k) \,
 a_\lambda (k) \,
\end{equation}
is the Hamiltonian for the free photon field, and the operator
\begin{equation}\nonumber
\begin{split}
  A(x) \; = \; \sum_{\lambda = +,-} \int_{\R^3} \,
  \frac{dk }{2 \pi |k|^{1/2}} \, \chL(|k|) \,
  \varepsilon_\lambda(k) \, \Big( \, e^{ikx} \otimes a_\lambda(k) \, + \,
  e^{-ikx} \otimes a_\lambda^\ast
  (k) \, \Big) \,   ,\\
\end{split}
\end{equation}
couples the electron to the quantized vector potential.
The polarization vectors $\varepsilon_\lambda(k)$, $\lambda\in\{+,-\}$, are unit vectors
orthogonal to one another and to $k\in\R^3$,
in accordance with the Coulomb gauge condition, ${\rm div}A =0$.

The $C^1$ function $\chL$ implements a fixed ultraviolet cutoff on the wavenumbers $k$.
Via scaling, we may assume $\chL$ to be compactly supported in $\{ |k|\leq 2\}$,
monotone, and to satisfy $\chL = 1$ for $|k|\leq  1$. For
convenience, we shall write
 $$
  A(x) \; = \; \Aa(x) \, + \, \Ac(x) ,
 $$
where
 $$
  \Aa(x) \; = \; \sum_{\lambda=+,-} \int_{\R^3}   \frac{dk }
  {2 \pi |k|^{1/2}} \, \chL(|k|) \, \varepsilon_\lambda(k) \, \mathrm{e}^{i k
  x} \, \otimes \, a_\lambda(k)
 $$
is the part of $A(x)$ containing the annihilation operators, and $\Ac(x)=(\Aa(x))^*$.


%
The system is translation invariant, and
$\PFop$ commutes with the operator of total momentum
 $$
 P_{tot} \; = \;  i \nabla_x \, \otimes \, \1_f \, + \, \1_{el} \, \otimes \, \Pf ,
 $$
where $i \nabla_x$ and
\eqn
    \Pf \; = \; \sum_{\lambda =+,-} \int dk \, k \, a^\ast_\lambda(k) \, a_\lambda(k)
\eeqn
denote the electron and the
photon momentum operators, respectively.
It therefore suffices to consider the restriction of $\PFop$ to the
fiber Hilbert space $\H_P \cong \Fo$
corresponding to the value $P\in\R^3$ of the conserved total momentum, given by
\begin{equation}
  \PFop(P) \; = \; \Hf \, + \, : \, ( \, P \, - \, \Pf \, - \sqrt{\alpha} \, A(0) \, )^2 \, : \; .
\end{equation}
Henceforth, we will write $A^\pm\equiv A^\pm(0)$. From \cite{BCV,Chen2006},
it is known that $\inf\sigma(\PFop)=\inf\sigma(\PFop(0))$, and we will in
the sequel only consider $P=0$.

\section{Statement of the Main Results}
On $\Fo$, we define the positive bilinear form
\begin{equation}\label{eq:def-scalar2}
 \la \, v \, , \, w \, \ra_* \; := \;
 \la \,  v \, , \, (\Hf + \Pf^2) \, w \, \ra ,
\end{equation}
and its associated semi-norm $\|v\|_* = \la v,v\ra_*^{1/2}$.

%
\begin{theorem}[Ground state energy of $\PFop(0)$]\label{thm:10-2}
We have
\eqn\label{eq:estimate-self-energy}
    \infspec(\PFop(0)) & = &\, - \, \alpha^2  \, \| \, \Phi_2 \, \|_*^2 \,
    \nonumber\\
    &&
    \, + \, \alpha^3 \, \big( \, 2 \, \|\Aa \Phi_2\|^2
    \, - \, 4 \, \|\Phi_3\|_*^2 \, - \, 4 \, \|\Phi_1\|_*^2 \, \big )
    \, + \, O(\alpha^{4})
\eeqn
where
\eqn\label{eq:E2}
    \Phi_2 &:=& - \,  (\Hf \, + \, \Pf^2 )^{-1} \, \Ac \cdot \Ac   \, \vac \, ,
    \\
    \Phi_3 &:=& - \,  (\Hf \, + \, \Pf^2 )^{-1} \, \Pf \cdot \Ac   \, \Phi_2 \, ,
    \label{eq:E3}
    \\
    \label{eq:E1}
    \Phi_1 &:=& - \,  (\Hf \, + \, \Pf^2 )^{-1} \, \Pf \cdot \Aa  \, \Phi_2 \, .
\eeqn
\end{theorem}

\begin{remark}
An easy computation shows that
\eqn
    \label{eq:Phij-norm-bd-1}
    C \; > \; \| \, \Phi_j \, \|_* \; > \; C^{-1}>0
\eeqn
for a constant $C<\infty$, and $j=1,2,3$.
\end{remark}


Let $\Psi \in\Fo$ be the minimizer of $\PFop(0)$, normalized by $\bra \Psi \, , \, \vac\ket=1$.
Taking the $\la \, \cdot \, , \,
\cdot \, \ra_*$-orthonormal projections of $\Psi$
along the vectors $\Phi_j$, $j=1,2,3$, and denoting the
component in the $\la \, \cdot \, , \,
\cdot \, \ra_*$-orthogonal complement of their span by $R$, we get
\eqn
    \label{eq:gs-exp-1}
    \Psi \; = \; \vac
    \, + \, 2 \, \eta_1 \, \alpha^{\frac32} \, \Phi_1
    \, + \, \eta_2 \, \alpha \, \Phi_2
    \, + \, 2 \, \eta_3 \, \alpha^{\frac32} \, \Phi_3
    \, + \, R
\eeqn
where for $j=1,2,3$
\eqn
    \label{eq:orthog-1}
    \bra \, \Phi_i \, , \, \Phi_j \, \ket_* \; = \;  \| \, \Phi_j \, \|_*^2 \; \delta_{ij}
\eeqn
\eqn
    \label{eq:orthog-2}
    \bra \, \Phi_j \, , \, R \, \ket_* \; = \; 0
    \; \; \; , \; \; \;
    \bra \, \vac \, , \, R \, \ket \; = \; 0 \; = \; \bra \, \vac \, , \, \Phi_j \, \ket \;.
\eeqn
The coefficients $\eta_j$ remain to be determined.


\begin{theorem}[Ground state of $\PFop(0)$]\label{thm:10-1}
Let $\Psi$ be the ground state (\ref{eq:gs-exp-1}) of $\PFop(0)$, normalized by
$\bra \Psi , \vac \ket =1$, \cite{BCFS2,Chen2006}. Then,
 \begin{equation}
 \begin{split}
  \Psi \; = \; \Omega_f \, + \,
  \alpha \, \Phi_2 \, + \, 2 \, \alpha^{\frac32} \,
  \Phi_1 \, + \, 2 \, \alpha^{\frac32} \, \Phi_3 \, + \widetilde R \,,
 \end{split}
 \end{equation}
where
\eqn
    \widetilde R \; := \; R \, + \, 2 \, (\eta_1-1) \, \alpha^{\frac32} \, \Phi_1
    \, + \, (\eta_2-1) \, \alpha \, \Phi_2
    \, + \, 2 \, (\eta_3-1) \, \alpha^{\frac32} \, \Phi_3 \,.
\eeqn
The coefficients  $\eta_j$
satisfy  $| \, \eta_{1,3}-1 \, | \, \leq \, c \, \alpha$, and $| \, \eta_{2}-1 \, | \, \leq \, c \, \alpha^2$.
Moreover,
\eqn
    \label{eq:R-bounds-1}
    \| \, \widetilde R \, \| \; , \; \| \, R \, \|  \; \leq \;  c \, \alpha
    \; \; \; \; , \; \mbox{and} \; \; \;
    \| \, \widetilde R \, \|_*  \; , \; \| \, R \, \|_* \; \leq \; c \,  \alpha^2 \, .
\eeqn

\end{theorem}

A key ingredient in our proof is the following estimate on the
expected photon number.

\begin{proposition}
\label{lm:Num-bd-1}
The expected photon number in the ground state is bounded by
\eqn
\label{eq:bound-photon}
 \bra \, \Psi \, , \,  N_f \, \Psi \, \ket \;
 \leq \; C \, \alpha^2 \, \bra \, \Psi \, , \, \Psi \, \ket
\eeqn
for a positive constant $C<\infty$ independent of $\alpha$, where
\eqn
 N_f \; = \; \sum_{\lambda=+,-} \int d k \, a^*_\lambda(k) \, a_\lambda(k)
\eeqn
is the photon number operator.
\end{proposition}

\begin{proof}
For $\sigma>0$, let $H_\sigma(P)$ denote the fiber Hamiltonian regularized by an infrared
cutoff implemented by replacing the function $\chL$ by a $C^1$ function
$\chL_\sigma$ with $\chL_\sigma=\chL$ on $[\sigma,\infty)$, $\chL_\sigma(0)=0$,
and $\chL_\sigma$ monotonically increasing on $[0,\sigma]$.
Then, $E_\sigma(P):=\infspec(H_\sigma(P))$ is a simple eigenvalue with
eigenvector $\Psi_\sigma(P)\in\Fo$, \cite{BCFS2,Chen2006}.
If $P=0$, one has $\nabla_P E_\sigma(P=0)=0$, \cite{BCFS2,Chen2006}.
In formula (6.11) of \cite{ChFr2006}, it is shown that
\eqn
    a_\lambda(k) \Psi_\sigma(0) \; = \; (I) \, + \, (II)
\eeqn
where from (6.12)
in \cite{ChFr2006} follows that
\eqn
    \| \, (I) \, \| \; \leq \; C(k) \, | \, \nabla_P E_\sigma(0) \, | \; = \; 0 \, ,
\eeqn
and
that
\eqn
    (II) \; = \; - \, \sqrt\alpha \, \frac{\chL_\sigma(|k|)}{|k|^{1/2}}
    \, \frac{1}{H_\sigma(k)-E_\sigma(0)+|k|}
    \, (H_\sigma(0)-E_\sigma(0)) \, \epsilon_\lambda(k) \cdot \nabla_P\Psi_\sigma(0)
\eeqn
if the electron spin is zero.
Thus, it follows immediately from (6.19) in \cite{ChFr2006} that
$$
    \| \, a_\lambda(k)\Psi_\sigma(0) \, \| \; \leq \;  c \, \sqrt\alpha \, \frac{\chL_\sigma(|k|)}{|k|} \,
    \Big| \, \frac{1}{m_{ren,\sigma}} \, - \, 1 \, \Big| \, \| \, \Psi_\sigma(0) \, \|
    \; \leq \;  c \, \alpha \, \frac{\chL_\sigma(|k|)}{|k|} \, \| \, \Psi_\sigma(0) \, \|  \,,
$$
for spin zero, where $m_{ren,\sigma}$ is the renormalized
electron mass for $P=0$, \cite{BCFS2,Chen2006}, defined by
\eqn
    \frac{1}{m_{ren,\sigma} } \; = \; 1 \, - \, 2 \,
    \frac{\bra \, \nabla_P \Psi_\sigma(0) \, , \, \big( \, H_\sigma(0) \, - \, E_\sigma(0) \, \big) \,
    \nabla_P \Psi_\sigma(0)  \, \ket }
    {\bra \,  \Psi_\sigma(0) \, , \,  \Psi_\sigma(0)  \, \ket} \,.
\eeqn
As proved in \cite{BCFS2,Chen2006}, $1 < m_{ren,\sigma} < 1 + c\alpha$ {\em uniformly} in $\sigma\geq0$.
Correspondingly, we obtain
$$
    \langle \, \Psi \, , \,  N_f \, \Psi \, \rangle
    \; = \; \lim_{\sigma\searrow0} \, \int d k \, \| \, a_\lambda(k)\Psi_\sigma(0) \, \|^2
    \; \leq \; C \, \alpha^2 \, \| \, \Psi \, \|^2
$$
as claimed, where $\Psi=s-\lim_{\sigma\searrow0}\Psi_\sigma(0)$ (see \cite{BCFS2}).
\end{proof}

\section{Proof of Theorems \ref{thm:10-2} and \ref{thm:10-1}}



To prove Theorems \ref{thm:10-2} and \ref{thm:10-1},
we derive the following upper and lower bounds
on $\infspec(\PFop(0))$.

\subsection{The upper bound}

We define the trial function
\begin{equation}\label{eq:cUtrial-def}
 \Psi^{\mathrm{trial}} \; := \; \vac \, + \, 2 \, \alpha^{\frac32} \Phi_1 \,  + \,
 \alpha \, \Phi_2
 \, + \, 2 \, \alpha^{\frac32}\Phi_3 \,  .
\end{equation}
Then, the variational upper bound on the ground state energy
\eqn
    \langle \,  \Psi^{\mathrm{trial}} \, , \, \PFop(0) \Psi^{\mathrm{trial}} \, \rangle
    & = &
    - \, \alpha^2 \, \|\Phi_2\|_*^2
    \label{eq:En-upbound-1}
    \\
    &&\, + \, \alpha^3
    \, \big( \, 2 \, \|\Aa \Phi_2\|^2 \, - \, 4 \, \|\Phi_3\|_*^2 \, - \, 4 \, \|\Phi_1\|_*^2 \, \big)
    \, + \, O(\alpha^{4}) \, ,
    \nonumber
\eeqn
is obtained from a straightforward computation.

\subsection{The lower bound}

Substituting the expression (\ref{eq:gs-exp-1}) for $\Psi$, and exploiting
$\bra \, \cdot \, , \, \cdot \, \ket_*$-orthogonality, straightforward
calculations explained in Section {\ref{sect:En-lowbound-proof}} yield
\eqn
    \bra \, \Psi \, , \, \PFop(0) \, \Psi \, \ket
    &\geq&
    \, - \, \alpha^2 \, \| \, \Phi_2 \, \|_*^2
    \, + \, \alpha^3 \, |\eta_2|^2 \, \big( \, 2  \, \| \, \Aa \, \Phi_2 \, \|^2
    \, - \, 4 \, \| \, \Phi_1 \, \|^2_* \, - \, 4 \, \| \, \Phi_3 \, \|^2_* \, \big)
    \nonumber\\
    &&
    \, + \, \frac12 \, \| \, R \, \|_*^2
    \nonumber\\
    &&
    \, + \, \alpha^2 \, |\eta_2 - 1 |^2 \, \| \, \Phi_2 \, \|_*^2 \,
    \nonumber\\
    &&
    \, + \, 4 \, \alpha^3 \, \big( \, |\eta_1-\eta_2|^2 \, \| \, \Phi_1 \, \|_*^2
    \, + \, |\eta_3-\eta_2|^2 \, \| \, \Phi_3 \, \|_*^2 \, \big)
    \nonumber\\
    &&
    \, - \, c \, \alpha^4 \, \big( \, 1 \, + \, |\eta_1|^2 \, + \, |\eta_2|^2 \, + \, |\eta_3|^2 \, \big) \,.
    \label{eq:En-lowbound-1}
\eeqn
We recall from (\ref{eq:Phij-norm-bd-1}) that $C>\| \, \Phi_j \, \|_*>C^{-1}>0$
for some finite $C$ independent of $\alpha$, and $j=1,2,3$.
We use
\eqn
    | \, \eta_2 \, - \, \eta_j \, |^2 \; \geq \; \frac12 \, | \, \eta_j-1 \, |^2 \, - \, | \, \eta_2-1 \, |^2
    \; \; \; \; \mbox{and} \; \; \; \;
    | \, \eta_j \, |^2 \; \leq \; 2 \, | \, \eta_j-1 \, |^2 \, + \, 2
\eeqn
to show that the three last lines in (\ref{eq:En-lowbound-1}) are bounded below by
\eqn
    \label{eq:En-lowbound-2}
    | \, \eta_1-1 \, |^2 \, \alpha^3 \, B_1
    \, + \, | \, \eta_2-1 \, |^2 \, \alpha^2 \, B_2
    \, + \, | \, \eta_2-1 \, |^2 \, \alpha^3 \, B_3
    \, - \, c \, \alpha^4 \, ,
\eeqn
where the constants $B_j>0$ are defined by
\eqn
    B_2&=& \| \, \Phi_2 \, \|_*^2
    \, - \, 4 \, \alpha \, (\| \, \Phi_1 \, \|_*^2 \, + \, \| \, \Phi_3 \, \|_*^2)
    \, - \, c \, \alpha^2
    \nonumber\\
    B_1&=&  2 \, \| \, \Phi_1 \, \|_*^2 \, - \, c \, \alpha
    \nonumber\\
    B_3 &=& 2 \, \| \, \Phi_3 \, \|_*^2 \, - \, c \, \alpha \, .
\eeqn
%
%
Clearly, $(\ref{eq:En-lowbound-2}) \, \geq \, - \, c\, \alpha^4$, and
the minimizing triple $(\eta_1,\eta_2,\eta_3)$ satisfies
\eqn
    | \, \eta_{1,3}-1 \, | \, \leq \, c \, \alpha
    \; \; \; , \; \mbox{and} \; \; \;
    | \, \eta_{2}-1 \, | \, \leq \, c \, \alpha^2 \, .
\eeqn
Therefore, (\ref{eq:En-lowbound-1}) is bounded from below by
\eqn
    - \, \alpha^2 \, \| \, \Phi_2 \, \|_*^2 \,
    \, + \, \alpha^3
    \, \big( \, 2 \, \|\Aa \Phi_2\|^2 \, - \, 4 \, \|\Phi_3\|_*^2 \, - \, 4 \, \|\Phi_1\|_*^2 \, \big)
    \, + \, \frac12 \, \| \, R \, \|_*^2
    \, - \, c \, \alpha^4 \, .
\eeqn
Combined with (\ref{eq:En-upbound-1}), we find $\| \, R \, \|_*  \, \leq \, c \, \alpha^2$.
This proves Theorems \ref{thm:10-2} and \ref{thm:10-1}.
\qed

\section{Proof of Inequality (\ref{eq:En-lowbound-1})}
\label{sect:En-lowbound-proof}

Clearly,
\eqn
  \bra \, \Psi \, , \, \PFop(0) \, \Psi \, \ket & = &
  \bra \, \Psi \, , \, ( \, \Hf \, + \, \Pf^2 \, ) \, \Psi \, \ket
  \label{eq:En-term-1}
  \\
  && \, + \, 4 \, \sqrt{\alpha} \, \Re \, \bra \, \Psi \, , \,  \Pf \, \Aa \, \Psi \, \ket
  \label{eq:En-term-2}
  \\
  && \, + \, 2 \, \alpha \, \Re \, \bra \, \Psi \, , \, \Aa \, \Aa \, \Psi \, \ket
  \label{eq:En-term-3}
  \\
  && \, + \, 2 \, \alpha \, \bra \, \Psi \, , \, \Ac \, \Aa \, \Psi \, \ket \, .
  \label{eq:En-term-4}
\eeqn
We estimate the quadratic form of $\PFop(0)$ on the true ground state $\Psi$.

\subsection{Some auxiliary estimates}
%
Since
\eqn
    \| \, \Aa \, \psi \, \| \; \leq \; c \, \| \, H_f^{\frac12}\, \psi \, \| \, ,
\eeqn
one has
\eqn
    \label{eq:Phij-normst-est-2}
    \| \, \Aa \, \psi \, \| \; , \; \| \, P_f \, \psi \, \|
    \; \leq \; c \,\| \, \psi \, \|_*
\eeqn
for all $\psi\in\Fo$ in the intersection of the domains of $H_f^{\frac12}$ and $P_f$.
Let for brevity
\eqn
    \widetilde\Psi \; := \; \vac
    \, + \, 2 \, \eta_1 \, \alpha^{\frac32} \, \Phi_1
    \, + \, \eta_2 \, \alpha \, \Phi_2
    \, + \, 2 \, \eta_3 \, \alpha^{\frac32} \, \Phi_3
\eeqn
so that $\Psi=\widetilde\Psi+R$.
We observe that from (\ref{eq:Phij-norm-bd-1}),
(\ref{eq:Phij-normst-est-2}), and the Schwarz inequality,
\eqn
    \label{eq:tildPhi-est-1}
    \| \, \Aa \, \widetilde\Psi \, \|^2 \; , \; \| \, P_f \, \widetilde \Psi \, \|^2
    \; \leq \; c \,   \alpha^2 \, |\eta_2|^2
    \, + \, c \, \alpha^3 \, \big( \, |\eta_1|^2 \, + \, |\eta_3|^2 \, \big) \, .
\eeqn
In addition, we recall that $A^\pm P_f=P_f A^\pm$ by the Coulomb gauge.

\subsection{The term (\ref{eq:En-term-1})}
First of all, the term (\ref{eq:En-term-1}) gives
\eqn
    \label{eq:En-term-1-1}
    \bra \, \Psi \, , \, ( \, H_f \, + \, P_f^2 \, ) \, \Psi \, \ket
    \; = \;
    \alpha^2 \, \| \, \Phi_2 \, \|_*^2 \, + \,
    4 \, \alpha^3 \, \big(\, \| \, \Phi_1 \, \|_*^2 \, + \, \| \, \Phi_3 \, \|_*^2 \, \big)
    \, + \, \| \, R \, \|_*^2
\eeqn
due to the orthogonality relations (\ref{eq:orthog-1}),
(\ref{eq:orthog-2}).

\subsection{The term (\ref{eq:En-term-2})}
We expand the term (\ref{eq:En-term-2}) according to $\Psi=\widetilde\Psi+R$.
Clearly,
\eqn
    \label{eq:En-term-2-1}
    4 \, \sqrt\alpha \,\Re \, \bra \, \widetilde \Psi \, , \,  P_f \,  \Aa \, \widetilde\Psi \, \ket
    & = & \, 8 \, \alpha^3 \, \Re \, \eta_1 \, \eta_2 \, \bra \, P_f \, \Phi_1\, , \, \Aa \, \Phi_2 \, \ket
    \nonumber\\
    & + & \,  8 \, \alpha^3 \, \Re \, \eta_2 \, \eta_3 \, \bra \, \Ac \, \Phi_2\, , \, P_f \, \Phi_3 \, \ket  \, .
\eeqn
Moreover, we have
\eqn
    4 \, \sqrt\alpha \,\Re \, \bra \, R \, , \,  P_f \,  \Aa \, R \, \ket
    \; \geq \; - \, \frac18 \, \| \, R \, \|_*^2 \,.
\eeqn
Here, we used
$\| \, \Aa \, R \, \| \, \leq \, c \, \| \, H_f^{\frac12}\, R \, \|$
combined with (\ref{eq:tildPhi-est-1}), and applied the Schwarz inequality in
the form $|AB| \, \leq \, \frac{\delta}{2} A^2 \, + \, \frac{1}{2\delta} B^2$ for any $0<\delta<\infty$.

Furthermore, we find
\eqn
    8 \, \sqrt\alpha \,\Re \, \bra \, \widetilde \Psi \, , \,  P_f \,  \Aa \, R \, \ket
    \; \geq \; - \, \frac18 \, \| \, R \, \|_*^2
    \, - \, c \, \alpha^4 \, (\eta_1^2 \, + \, \eta_3^2)
\eeqn
by similar arguments, using
\eqn
    \bra \, \Phi_2 \, , \,  P_f \,  \Aa \, R \, \ket \; = \; \bra \, \Phi_3 \, , \,  R \, \ket_*
    \; = \; 0 \; = \; \bra \, \vac \, , \,  P_f \,  \Aa \, R \, \ket
\eeqn
(see (\ref{eq:orthog-2})), combined with (\ref{eq:tildPhi-est-1}).

\subsection{The term (\ref{eq:En-term-3})}
We have
\eqn
    2 \, \alpha \, \Re \, \bra \, \Psi \, , \, \Aa \, \Aa \, \Psi \, \ket
    \; = \;
    2 \, \alpha \, \Re \, \bra \, \Ac \, \Psi \, , \, \Aa \, \Psi \, \ket \,.
    \nonumber
\eeqn
Clearly,
\eqn
    \| \, \Ac \, \psi \, \|^2 \; = \; [\Aa,\Ac] \, \| \, \psi \, \|^2 \, + \, \| \, \Aa \, \psi \, \|^2 \,,
\eeqn
where $0<[\Aa,\Ac]<c$, for all $\psi\in\Fo$.
Using the Schwarz inequality and similar arguments as above, one finds
\eqn
    \label{eq:En-term-4-1}
    2 \, \alpha \, \Re \, \bra \, \Psi \, , \, \Aa \, \Aa \, \Psi \, \ket&\geq&
    2 \,  \alpha^2 \, \Re \, \eta_2 \, \bra \, \Ac \, \Ac \, \vac \, , \, \Phi_2 \, \ket
    \nonumber\\
    &&
    \, - \, \frac18 \, \| \, R \, \|_*^2 \, - \, c \, \alpha^2 \, \| \, R \, \|^2
    \nonumber\\
    &&
    \, - \, c \, \alpha^4 \, \big( \, |\eta_1|^2 \, + \, |\eta_2|^2 \, + \, |\eta_3|^2 \, \big) \,,
\eeqn
where we used that if $\psi=\vac$, $\Phi_1$, $\Phi_3$, or $R$, then by (\ref{eq:orthog-1}), (\ref{eq:orthog-2}),
\eqn
    \bra \, \Ac \, \Ac \, \vac \, , \, \psi \, \ket \; = \; \bra \, \Phi_2 \, , \, \psi \, \ket_* \; = \; 0 \,.
\eeqn
At this point, we apply Proposition {\ref{lm:Num-bd-1}} in
\eqn
    \| \, R \, \|  \; \leq \; \| \, N_f^{\frac12} \, R \, \|
    \; \leq \; \Big| \, \| \, N_f^{\frac12} \, \Psi \, \|
    \, - \, \| \, N_f^{\frac12} \, \widetilde\Psi \, \| \, \Big| \; \leq \; c \, \alpha \, .
\eeqn
Here, we use the fact that $R$ has a vanishing projection on the Fock vacuum, (\ref{eq:orthog-2}),
and $\| \, N_f^{\frac12} \, \widetilde\Psi \, \| \, \leq \, c \, \alpha$,
which one easily verifies.

\subsection{The term (\ref{eq:En-term-4})}
This term is estimated by
\eqn
    2 \, \alpha \, \| \, \Aa \, \Psi \, \|^2 \; \geq \; 2 \, \alpha^3 \, \| \, \Aa \, \Phi_2 \, \|^2
    \, - \, c \, \alpha^4 \, \big( \, |\eta_1|^2 \, + \, |\eta_3|^4 \, \big)
    \, - \, \frac18 \, \| \, R \, \|_*^2 \, .
\eeqn

\subsection{Collecting all estimates}
Combining the $O(\alpha^3)$-terms of (\ref{eq:En-term-1-1}) with (\ref{eq:En-term-2-1}), we find
\eqn
    \lefteqn{
    4 \, \alpha^3 \, |\eta_1|^2 \, \| \, \Phi_1 \, \|_*^2
    \, + \, 8 \, \alpha^3 \, \Re \, \eta_1 \, \eta_2 \, \bra \, \Phi_1\, , \, P_f \, \Aa \, \Phi_2 \, \ket
    }
    \nonumber\\
    & = & - \, 4 \, \alpha^3 \, |\eta_2|^2 \, \| \, \Phi_1 \, \|_*^2
    \, + \, 4 \, \alpha^3 \, |\eta_1-\eta_2|^2 \, \| \, \Phi_1 \, \|_*^2 \,,
\eeqn
since from (\ref{eq:E1}),
\eqn
    \bra \, \Phi_1\, , \, P_f \, \Aa \, \Phi_2 \, \ket \; = \; - \, \| \, \Phi_1 \, \|_*^2 \, .
\eeqn
Likewise,
\eqn
    \lefteqn{
    4 \, \alpha^3 \, |\eta_3| \, \| \, \Phi_3 \, \|_*^2
    \, + \, 8 \, \alpha^3 \, \Re \, \eta_2 \, \eta_3 \, \bra \, P_f \, \Ac \, \Phi_2\, , \, \Phi_3 \, \ket
    }
    \nonumber\\
    & = & - \, 4 \, \alpha^3 \, |\eta_2|^2 \, \| \, \Phi_3 \, \|_*^2
    \, + \, 4 \, \alpha^3 \, |\eta_2-\eta_3|^2 \, \| \, \Phi_3 \, \|_*^2 \,,
\eeqn
since from (\ref{eq:E3}),
\eqn
    \bra \, P_f \, \Ac \, \Phi_2\, , \, \Phi_3 \, \ket \; = \; - \, \| \, \Phi_3 \, \|_*^2 \, .
\eeqn
Combining the $O(\alpha^2)$-term in (\ref{eq:En-term-1-1}) with the first term
on the r.h.s. of (\ref{eq:En-term-4-1}),
\eqn
    \lefteqn{
    \alpha^2 \, |\eta_2|^2 \, \| \, \Phi_2 \, \|_*^2
    \, + \, 2 \, \alpha^2 \, \Re \, \eta_2 \, \bra \, \Ac \, \Ac \, \vac \, , \, \Phi_2 \, \ket
    }
    \nonumber\\
    & = & - \, \alpha^2 \, \| \, \Phi_2 \, \|_*^2
    \, + \, \alpha^2 \, |\eta_2 - 1 |^2 \,
    \| \, \Phi_2 \, \|_*^2 \, ,
\eeqn
since from (\ref{eq:E2}),
\eqn
    \bra \, \Ac \, \Ac \, \vac \, , \, \Phi_2 \, \ket \; = \; - \, \| \, \Phi_2 \, \|_*^2 \, .
\eeqn
Moreover, we use half of $\|R\|_*^2$ in  (\ref{eq:En-term-1-1}) to compensate all four
terms $-\frac18\| \, R \, \|_*^2$ appearing in the above estimates, and are left with
the term $\frac12\| \, R \, \|_*^2$ in (\ref{eq:En-lowbound-1}).

All remaining terms are bounded below by $- \, c \, \alpha^4 \,
\big( \, 1 \, + \, |\eta_1|^2 \, + \, |\eta_2|^2 \, + \, |\eta_3|^2 \, \big)$.

Collecting all bounds, we arrive at (\ref{eq:En-lowbound-1}).
\qed

\section*{Acknowledgments}

The authors gratefully acknowledge financial support from the
following institutions: The European Union through the IHP network
``Analysis and Quantum'' HPRN-CT-2002-00277 (J.-M. B. and S. V.), the French
Ministry of Research through the ACI ``jeunes chercheurs" (J.-M. B.), and
the DFG grant WE 1964/2 (S. V.). T. C. was supported by NSF Grant
DMS-0524909. V. V. thanks I.~M. Sigal for partial support by NSF
Grant DMS-0400526 and by NSERC Grant NA7901.

\end{document}